\documentclass[prl,aps,twocolumn]{revtex4-1}
\usepackage{mathbbold}
\usepackage{amsfonts}
\usepackage{amsmath}
\usepackage{graphicx}
\usepackage{subfig}
\begin{document}
\title{Likelihood transform: making optimization and parameter estimation easier}
\author{Yan Wang}
\affiliation{Max-Planck-Institut f\"ur Gravitationsphysik (Albert-Einstein-Institut), Callinstra{\ss}e 38, 30167 Hannover, Germany}
\email{yan.wang@aei.mpg.de} \pacs{}

\begin{abstract}
Parameterized optimization and parameter estimation is of great
importance in almost every branch of modern science, technology and engineering.
A practical issue in the problem is that when the parameter space is large and the available data is
noisy, the geometry of the likelihood surface in the parameter space
will be complicated. This makes searching and optimization
algorithms computationally expensive, sometimes even beyond reach. In this paper, we define a
likelihood transform which can make the structure of the likelihood
surface much simpler, hence reducing the intrinsic complexity and
easing optimization significantly. We demonstrate the properties of likelihood transform
by apply it to a simplified gravitational wave chirp signal search.
For the signal with an signal-to-noise ratio 20, likelihood transform has made
a deterministic template-based search possible for the first time, which
turns out to be 1000 times more efficient than an exhaustive grid-based search.
The method in principle can be applied to other problems in other fields as the
spirit of parameterized optimization and parameter estimation problem is the same.
\end{abstract}
\maketitle
%%%%%%%%%%%%%%%%%%%%%%%%%%%%%%%%%%%%%%%%%%%%%%%%%%%%%%%%%%

\emph{Introduction}--Parameterized optimization and parameter estimation is a general
important problem in almost every branch of modern science, technology and engineering \cite{Poisson95,Tamura11,Bollerslev92,Schmidt86,Johansen90}. The general problem
can be described as follows. Let us denote $\bbtheta=(\theta^1,\theta^2,...,\theta^k)\in \mathcal{P}$ as the parameters to be estimated,
where $\mathcal{P}$ is the $k$-dimensional parameter space.
The figure of merit $\mathcal{F}(\bbtheta,\mathbf{x})\equiv\mathcal
{M}\cdot(\bbtheta,\mathbf{x})$ is usually some functional of the
parameters $\bbtheta$ and the data $\mathbf{x}$ (e.g. measurement
data from experiments). In practice, the functional $\mathcal{M}$ can be viewed as a set of operations on
the parameters $\bbtheta$ and the data $\mathbf{x}$. The goal is to find the best
estimate $\bbtheta_*$ that maximizes the functional $\mathcal{F}(\bbtheta_*,\mathbf{x})$. Since minimization can be cast into maximization by just adding a minus sign,
we will focus only on maximization problems. Also, for brevity,
we omit $\mathbf{x}$ and denote $\mathcal{F}(\cdot,\mathbf{x})$ as $\mathcal{F}(\cdot)$ from now
on. The mapping $\mathcal{M}\cdot(\cdot,\mathbf{x})$ from
$\bbtheta$ to $\mathcal{F}$ defines a hypersurface in the
$k$-dimensional parameter space $\mathcal{P}$. Hereafter, we refer
it as the \emph{likelihood surface} in general, although sometimes it does not have to be related to the likelihood.
%, which is a key element in signal detection and parameter estimation.

By definition, the likelihood surface should peak at the best estimate $\bbtheta_*$. When the peak is broad and smooth,
there are less structures in this region. Hence, the search is relatively easy and the resolution is poor (i.e. the error bar
in the estimate is large). When the peak is sharp and narrow, the resolution is high, but the search is much more difficult.
%For grid-based search, the resolution of the grid should be better than the size of the peak. For heuristic search, the peak region
%is very tiny compared to the entire parameter space.
In general, the structure of the likelihood surface determines the difficulty of the optimization problem. By modifying the surface structure, we may alter the innate difficulty of the problem.

\emph{Likelihood transform}--We introduce a set of functionals $\mathcal{Y}_\bbsigma$ acting on the mapping functional $\mathcal{M}$,
where $\bbsigma\in \mathcal{Q}$ can be either a scalar variable or a set of variables.
\begin{eqnarray}
\mathcal{F}_\bbsigma(\cdot)&\equiv&\mathcal{Y}_\bbsigma\cdot[\mathcal{M}\cdot(\cdot,\mathbf{x})]\nonumber\\
&=&(\mathcal{Y}_\bbsigma\cdot\mathcal{M})\cdot(\cdot,\mathbf{x}),
\end{eqnarray}
By varying $\bbsigma$, we obtain a set of modified likelihood surfaces $\mathcal{F}_\bbsigma(\cdot)$. We want to
find a proper set of functionals $\mathcal{Y}_{\bbsigma(l)}$, where the index $l\in [0,l_*]\subset \mathbb{R}$, such that as $l$ running from $l_*$ to $0$, $\mathcal{Y}_{\bbsigma(l)}$ modifies the sharp narrow peak at the best estimate $\bbtheta_*$ gradually (or continuously)
to broader and smoother hills. We require that $\mathcal{Y}_{\bbsigma(l_*)}$ is a unity functional, i.e. $\mathcal{F}_{\bbsigma(l_*)}(\cdot)=\mathcal{F}(\cdot)$.
When such a proper set of functionals is identified, one can search on the broadest and smoothest likelihood surface, $\mathcal{F}_{\bbsigma(0)}(\cdot)$ , since its (local) maximum $\bbtheta_\bbsigma$ is usually easiest to find. Notice that
this maximum $\bbtheta_\bbsigma$ needs not to be the global maximum on the modified likelihood surface and it needs not to be exactly at $\bbtheta_*$. However, as $l$ going from $0$ to $l_*$, $\bbtheta_\bbsigma$ should gradually
converge to $\bbtheta_*$ due to the continuity of the transform. This means after identifying the maximum $\bbtheta_{\bbsigma(0)}$ in the smoothest likelihood surface,
the transform $\mathcal{Y}_{\bbsigma(l)}$ can help lead us to the best estimate $\bbtheta_*$.

Following the conventions \cite{Jaranowski12} used by the gravitational wave (GW) data analysis community, it is convenient to define the inner product of two time
series $a(t),b(t)$ as below
\begin{eqnarray}
\langle a|b\rangle=\int_{-\infty}^{\infty}\frac{\tilde{a}^*(f)\tilde{b}(f)}{S_n(f)}df,
\end{eqnarray}
where $\tilde{a}(f),\tilde{b}(f)$ are the Fourier transforms of $a(t)$ and $b(t)$. $S_n(f)$ is the so-called two-sided
power spectral density of Gaussian noise, usually defined as
$E[\tilde{n}^*(f')\tilde{n}(f)]=S_n(f)\delta(f-f')$.

We denote the normalized GW waveform with parameters $\bbtheta$ by $h(\bbtheta,t)$, thus $\langle h(\bbtheta)|h(\bbtheta) \rangle=1$. The measured data
$x(t)$ containing a GW signal with parameters $\bbtheta_*$ and Gaussian noise $n(t)$ can be expressed as $x(t)=Ah(\bbtheta_*)+n(t)$, where $A$ is the strength of the signal.

The figure of merit is the signal-to-noise ratio (SNR)
\begin{eqnarray}\label{MatchFilter}
\mathcal{F}(\bbtheta)&=&\mathcal{M}\cdot (\bbtheta,\mathbf{x})\nonumber\\
&\equiv&\langle x|h(\bbtheta) \rangle \nonumber\\
&=&\int_{-\infty}^{\infty}\frac{\tilde{x}^*(f)\tilde{h}(\bbtheta,f)}{S_n(f)}df.
\end{eqnarray}
Although $\mathcal{F}(\bbtheta)$ here is not the likelihood, it is directly related to the likelihood $\mathcal{L}(\bbtheta)\propto \exp[\mathcal{F}(\bbtheta)^2/2]$.
The functional $\mathcal{M}$ can be interpreted as two operations: first, to generate a waveform with parameters $\bbtheta$; second, to calculate the inner product of this waveform and the data $\mathbf{x}$. Usually, $\mathcal{F}(\cdot)$ has a shape narrow peak at the best estimate $\bbtheta_*$. As an example, we define the functionals $\mathcal{Y}_{\bbsigma}$ as convolution operators
\begin{eqnarray}\label{Eq:Kh}
\mathcal{F}_{\bbsigma}(\bbtheta) &=& (\mathcal{Y}_{\bbsigma}\cdot\mathcal{M})\cdot (\bbtheta,\mathbf{x})\nonumber\\
&\equiv&\langle x| \mathcal{K}_{\bbsigma} \star h(\bbtheta) \rangle \nonumber\\
&=& \mathcal{K}_{\bbsigma} \star \mathcal{F}(\bbtheta),
\end{eqnarray}
where $\mathcal{K}_{\bbsigma}$ is the kernel function. The last equality is because convolution is a linear operation and $\mathcal{F}(\bbtheta)$ is linear in $h(\bbtheta)$. Since the convolution can be viewed as a smoothing functional, the modified likelihood surface $\mathcal{F}_{\bbsigma}(\cdot)$ is smoother than the original one. For brevity's sake, we temporarily assume $\bbtheta$ is a scaler parameter and choose the kernel function as one-dimensional Gaussian function $\mathcal{K}_{\bbsigma}=\frac{1}{\sqrt{2\pi}\bbsigma}\exp(-\frac{\bbtheta^2}{2\bbsigma^2})$. The argument below can be trivially generalized to multi-dimensional case. When $\bbsigma$ is large, the kernel is a very broad Gaussian function, hence making the likelihood surface $\mathcal{F}_{\bbsigma}(\cdot)$ very smooth. As $\bbsigma$ decays to $0$, $\mathcal{F}_{\bbsigma}(\cdot)$ converges to the original likelihood surface $\mathcal{F}(\cdot)$. Notice that when $\bbsigma\rightarrow 0$, we have $\mathcal{K}_{\bbsigma}\rightarrow \delta(\bbtheta)$. In practice, we can set $\bbsigma(l)=\bbsigma(0)(l_*-l)/l_*$. As $l$ goes from $0$ to $l_*$, $\mathcal{F}_{\bbsigma(l)}(\cdot)$ evolves gradually from very smooth modified likelihood surface to the original likelihood surface.

From another point of view, $\mathcal{K}_{\bbsigma} \star h(\bbtheta)$ in Eq.~\ref{Eq:Kh} is just a weighted average of many waveforms. Since waveforms with similar parameters are correlated, by using a summation of nearby waveforms one can smooth the original likelihood surface. As the number of averaged waveforms goes to 1, the modified likelihood surface converges to the original likelihood surface.

\emph{Applications}--Likelihood transform $\mathcal{Y}_{\bbsigma(l)}$ can gradually modify the likelihood surface, hence changing the intrinsic complexity
of the optimization problem. In the meantime, it retains the relation between the modified likelihood surfaces and the original likelihood surface.
Therefore, likelihood transform can be used in many ways. For instance, it can accelerate stochastic optimization methods, such as Markov chain Monte Carlo~\cite{Littenberg:2009bm,Wang12}, particle swarm optimization~\cite{Wang10,Wang12}, genetic algorithm~\cite{Crowder:2006wh,Petiteau:2010zu,Wang12}, etc.
It can help design hierarchical search algorithms. In some circumstances, it can even make a deterministic search possible.

\begin{figure}[htbp]
\subfloat[Simulated Annealing]{
\begin{minipage}[t]{0.25\textwidth}
\centering
\includegraphics[width=1.0\textwidth]{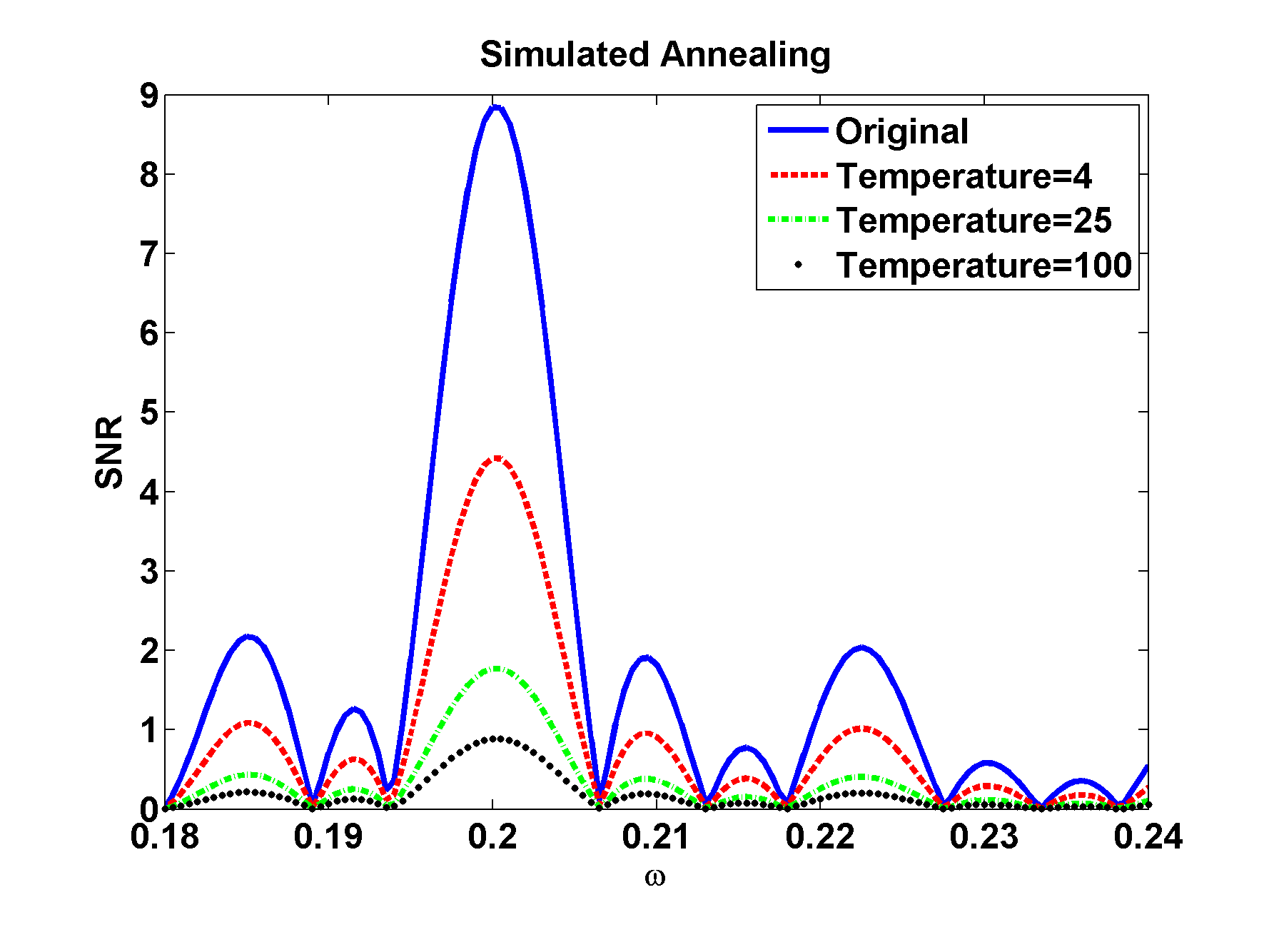}
\end{minipage}
}
\subfloat[Likelihood Transform]{
\begin{minipage}[t]{0.25\textwidth}
\centering
\includegraphics[width=1.0\textwidth]{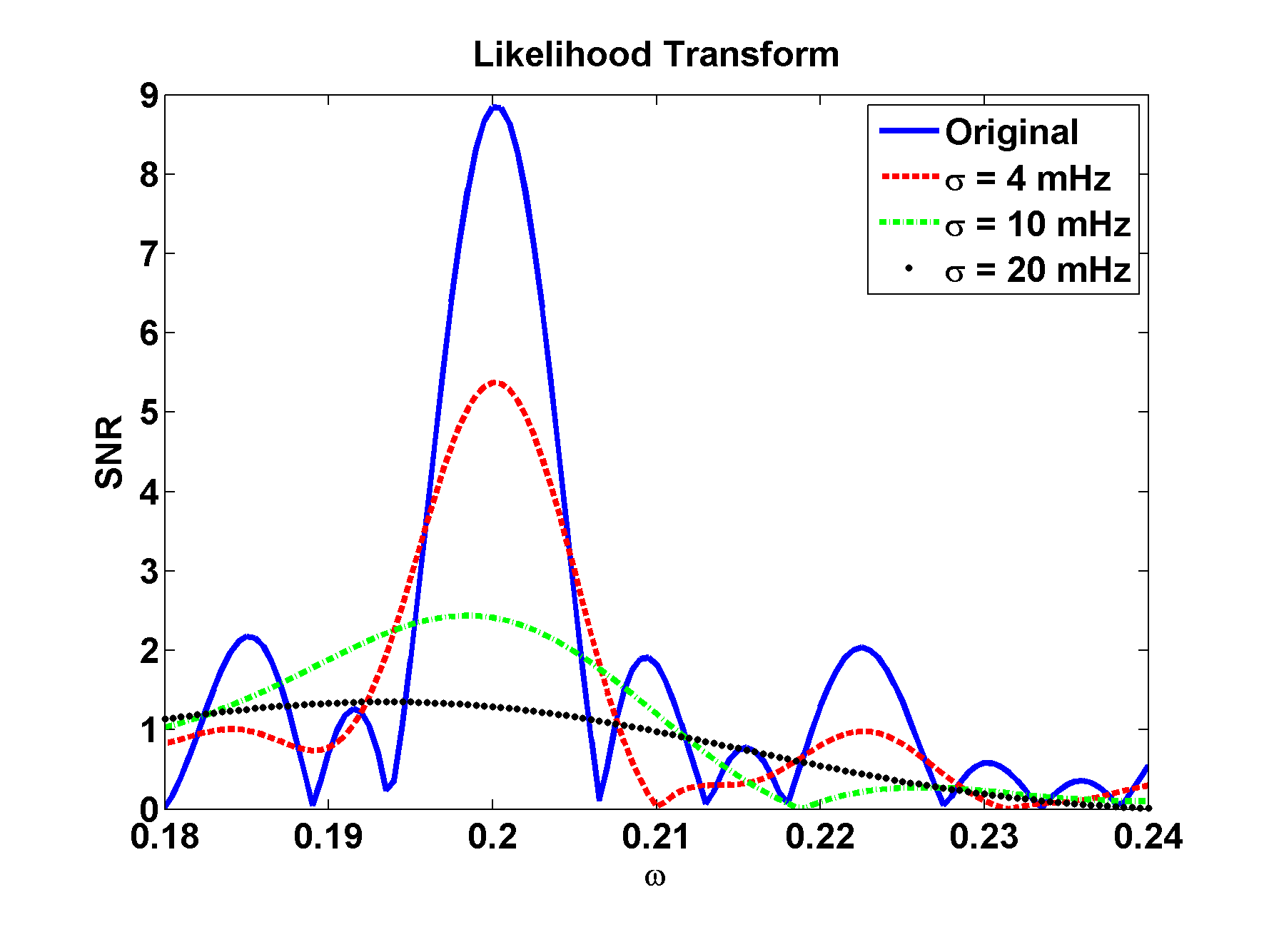}
\end{minipage}
}
\caption{ \label{fig:SAvsLT} A comparison of simulated annealing and likelihood transform. }
\end{figure}

Likelihood transform is different from simulated annealing \cite{Kirkpatrick83}, which is also a technique to accelerate stochastic optimization algorithms. In the following, we will compare the two. Simulated annealing employs a temperature parameter $\mathcal{T}$ to heat the likelihood surface from
$\mathcal{L}(\bbtheta)\propto \exp[ \mathcal{F}(\bbtheta)^2/2]$ to $\exp[ \mathcal{F}(\bbtheta)^2/2\mathcal{T}]$. As the stochastic
search algorithms proceed, the temperature cools down gradually. Therefore, the heuristics can escape from local maxima more easily and explore the whole
parameter space more thoroughly, hence being accelerated. As an example, we simulated a sinusoidal signal with only one parameter $\omega=0.2\,$rad/s buried
in white Gaussian noise. The SNR was 9. Fig.~\ref{fig:SAvsLT} (a) shows how simulated annealing gradually modifies the likelihood surface (or more rigorously
the SNR surface). As seen from the figure, the likelihood surface is less spiky at high temperatures. Notice that the number of local maxima (including the global
maximum) is the same at all temperatures, and the locations of these maxima are unchanged.

As for likelihood transform, it modifies the likelihood as $\exp[ \mathcal{F}_{\bbsigma (l)}(\bbtheta)^2/2]$. For simplicity, we choose the functional $\mathcal{Y}_\bbsigma$ to be convolution operators with a Gaussian kernel. We then applies it to the same simulated data. The modified likelihood surface
$\mathcal{F}_{\bbsigma (l)}(\bbtheta)$ are shown in Fig.~\ref{fig:SAvsLT} (b). Notice that both the number of local maxima and their locations are changed
by the likelihood transform. In addition, the likelihood surfaces are smoother with less structures comparing to the cases of simulated annealing. These
features of likelihood transform may help accelerate the stochastic optimization algorithms more efficiently.

Next, we applied likelihood transform to a toy model in
gravitational wave data analysis. Although it is a somehow simplified
model, it captures important features of the general problem and
can demonstrate the method in a more general (less
background-dependent) way.

Suppose the normalized gravitational wave chirp signal is in the following form
\begin{eqnarray}
h(\bbtheta,t)=\sqrt{\frac{2}{T}}\cos[\omega_0 \alpha_0(t)+\omega_1 \alpha_1(t) ],
\end{eqnarray}
where $\bbtheta=(\omega_0,\omega_1)$ are the two parameters to be estimated, the two time functions are defined as $\alpha_0(t)\equiv t-\frac{T}{2}$ and $\alpha_1(t)\equiv (t-\frac{T}{2})^2$, $T$ is the total observation time, which we choose to be $5120$ seconds in the simulation.
We inject a signal with parameters
$\omega_0=0.0628\,\textrm{rad/s},\,\omega_1=6.136\times10^{-6}\,\textrm{rad/s}^2$.
Notice that $\omega_0$ is an angular frequency. The searching parameter
ranges for $\omega_0$ and $\omega_1$ are
$(\omega_0^\textrm{low},\omega_0^\textrm{upp})=(1.2\,\textrm{mrad/s},0.126\,\textrm{rad/s})$ and
$(\omega_1^\textrm{low},\omega_1^\textrm{upp})=(-3.07\times10^{-6}\,\textrm{rad/s}^2,1.23\times10^{-5}\,\textrm{rad/s}^2)$
respectively.

\begin{figure}
\includegraphics[width=0.5\textwidth]{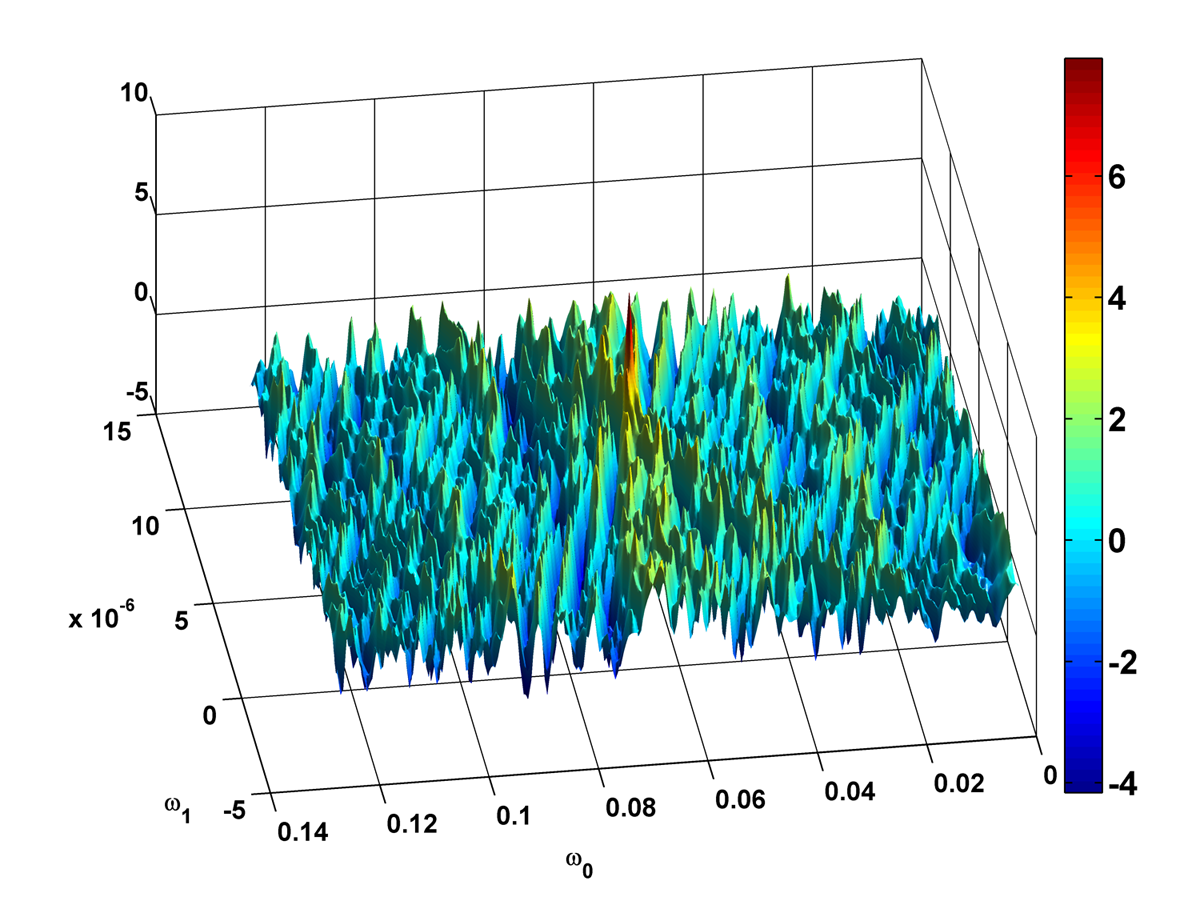}
\caption{ \label{fig:likelihood0} The original likelihood surface
$\mathcal{F}(\omega_0,\omega_1)$. It peaks at the true signal
parameter with an optimal SNR 8. It is very fluctuant. Optimal
template layout requires 69620 templates.}
\end{figure}

As an example, we use convolution operators as the functionals $\mathcal{Y}_\bbsigma (l)$ and assume the kernel function to be a Gaussian function with
diagonal covariance. Then, we have
\begin{eqnarray}
H_{\bbsigma (l) } (\bbtheta) &=&  \mathcal{K}_{\bbsigma (l)} \star h(\bbtheta) , \nonumber \\
&=& \int \int \mathcal{K}[\omega_0-\omega_0',\omega_1-\omega_1',\sigma_0(l),\sigma_1(l)]  \nonumber \\
 && \cdot \, h(\omega_0',\omega_1') \textrm{d} \omega_0' \textrm{d} \omega_1' , \nonumber \\
&=& h(\bbtheta) e^{ -\frac{1}{2} [  \sigma_0(l)^2 \alpha_0(t)^2 +  \sigma_1(l)^2 \alpha_1(t)^2  ] }, \\
\mathcal{F}_{\bbsigma (l)}(\bbtheta) &=& \langle x| H_{\bbsigma (l) } (\bbtheta) \rangle .
\end{eqnarray}
We set $l_*=1$ and choose $(\sigma_0(l),\sigma_1(l)) = (1-l)(\omega_0^{upp}-\omega_0^{low},\omega_1^{upp}-\omega_1^{low})$
to be a fraction of the entire searching parameter range.
We will see how this parameter $l$ can modify the
likelihood surface and adjust the difficulty of the optimization
problem. In general, the difficulty of the search can be very well described by
the required number of templates for a certain mismatch by
template-based search. Following conventions, we set the
mismatch to be $0.03$. By calculating the
metric of the likelihood surface on the parameter space \cite{Owen96}, it's
straightforward to estimate the number of templates required by
optimal layout (we choose rectangular layout here). When $l=l_*=1$, we
have the original likelihood surface
$\mathcal{F}(\omega_0,\omega_1)$ shown in Fig.~\ref{fig:likelihood0}.
On this likelihood surface, the optimal layout requires 69,620
templates. This large required number is due to the noise-like features of
the likelihood surface.

\begin{figure}[htbp]
\subfloat[$1-l=1/4$]{
\begin{minipage}[t]{0.25\textwidth}
\centering
\includegraphics[width=1.0\textwidth]{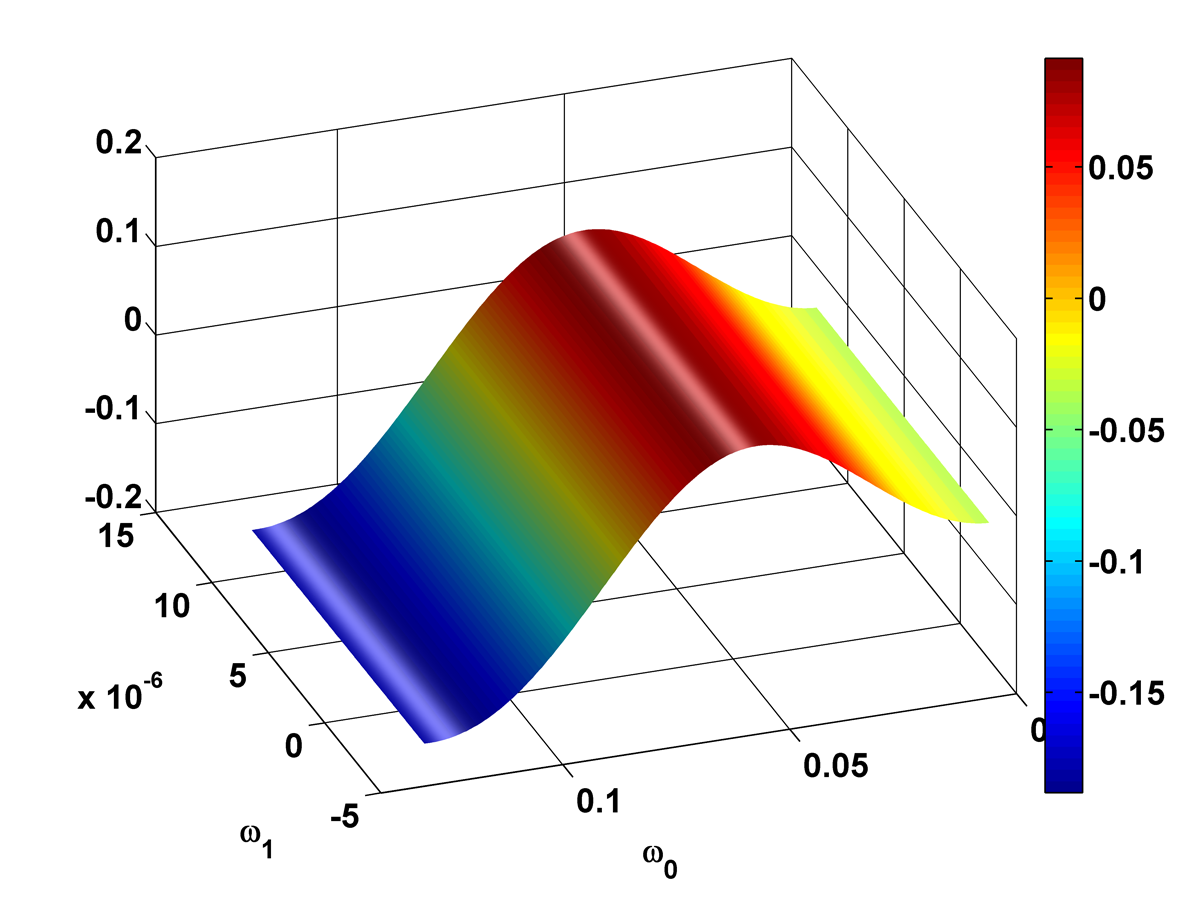}
\end{minipage}
}
\subfloat[$1-l=1/16$]{
\begin{minipage}[t]{0.25\textwidth}
\centering
\includegraphics[width=1.0\textwidth]{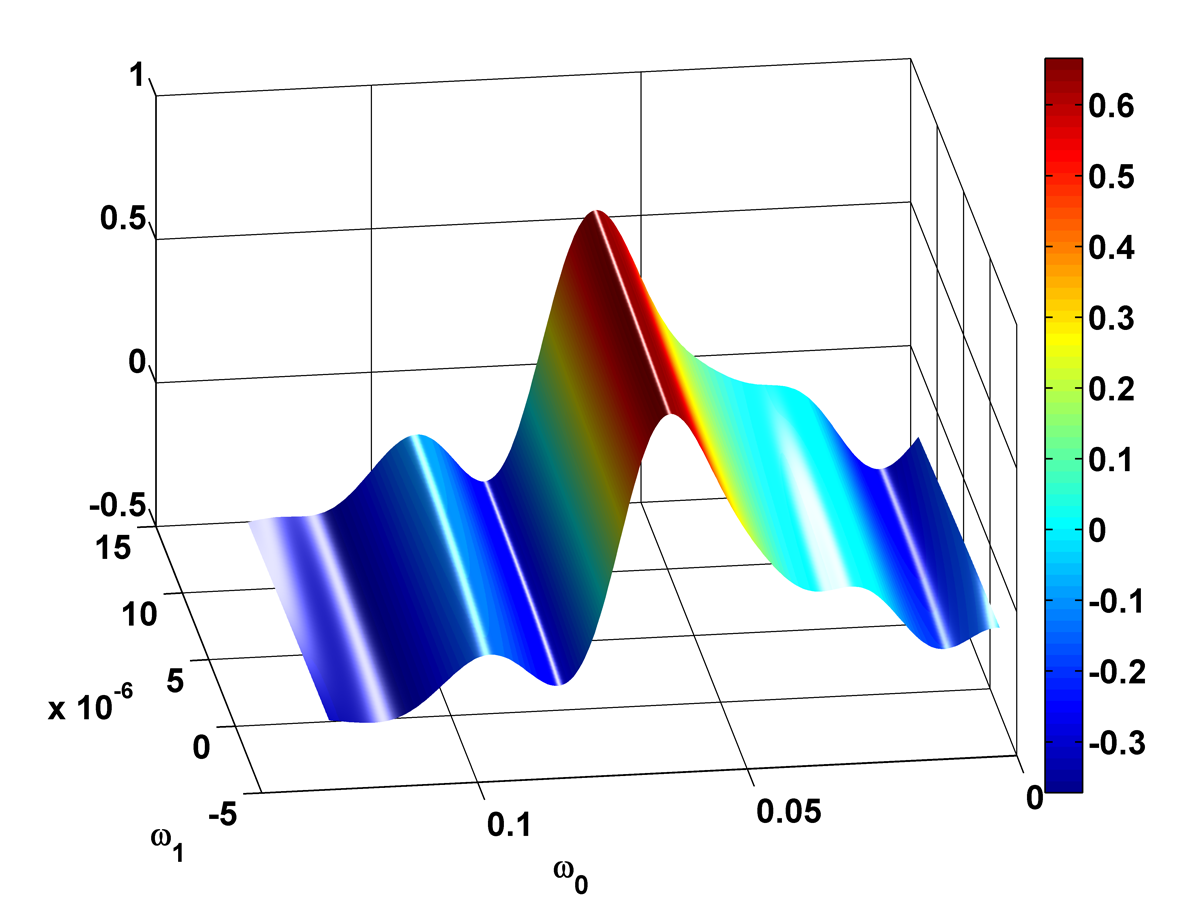}
\end{minipage}
}
\\
\subfloat[$1-l=1/64$]{
\begin{minipage}[t]{0.25\textwidth}
\centering
\includegraphics[width=1.0\textwidth]{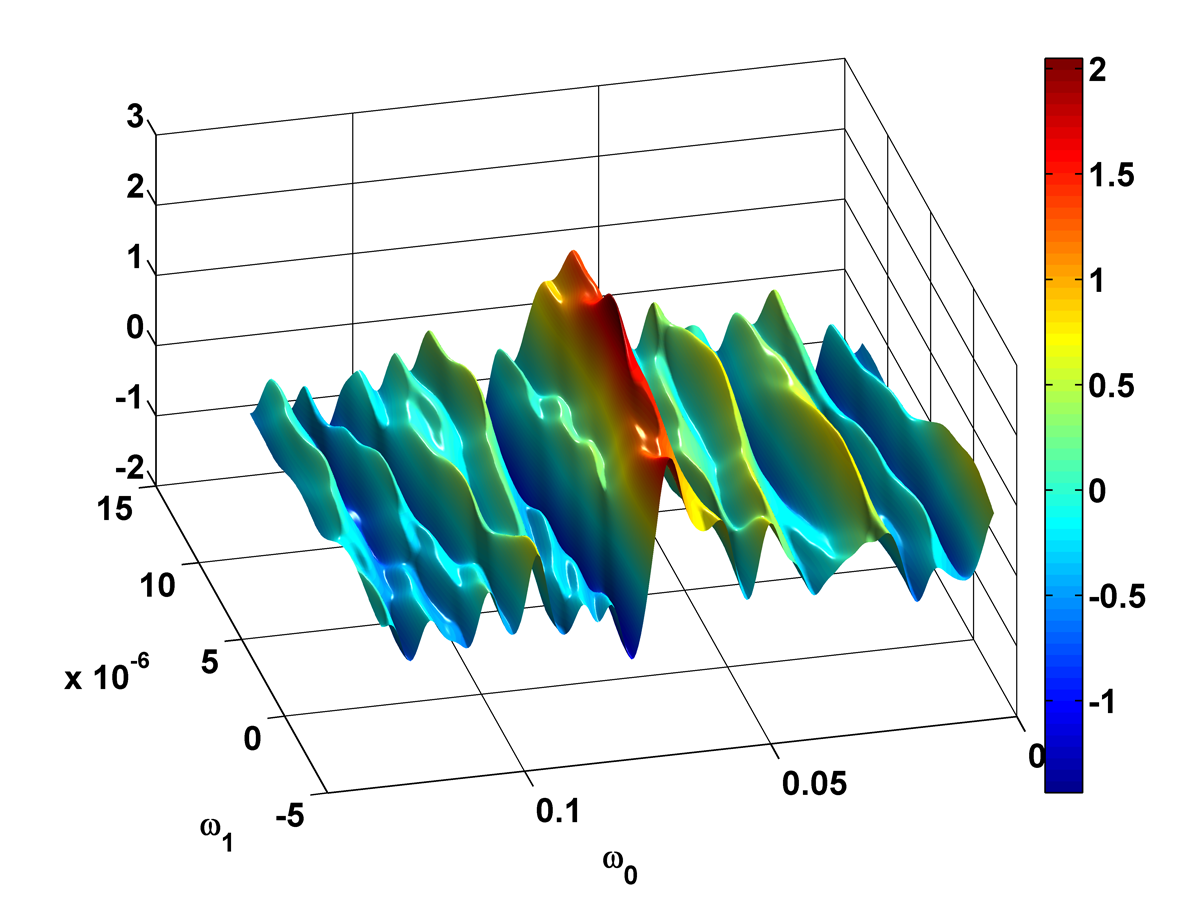}
\end{minipage}
}
\subfloat[$1-l=1/128$]{
\begin{minipage}[t]{0.25\textwidth}
\centering
\includegraphics[width=1.0\textwidth]{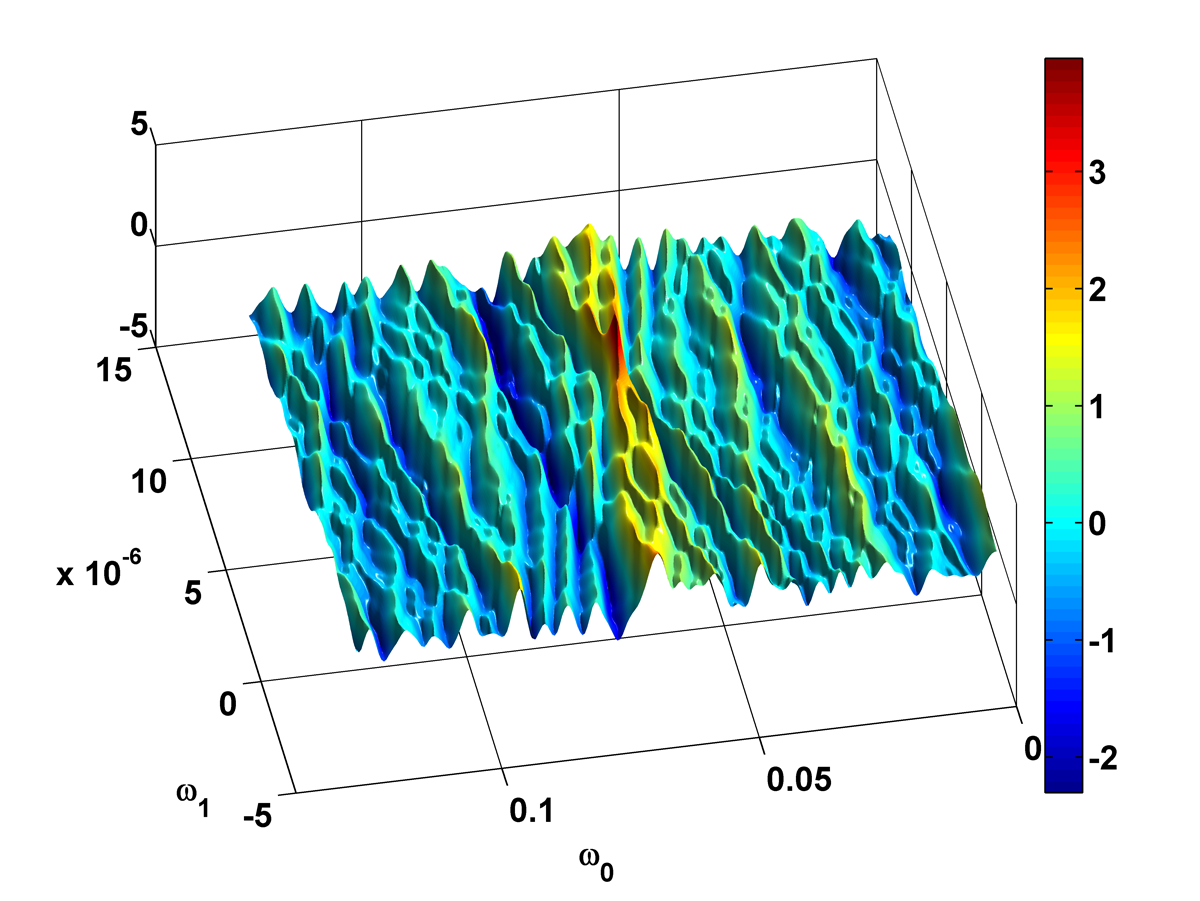}
\end{minipage}
}
\caption{ \label{fig:likelihood3578} The modified likelihood surfaces $\mathcal{F}_{\bbsigma(l)}(\omega_0,\omega_1)$ after likelihood transforms. }
\end{figure}

\begin{figure}
\includegraphics[width=0.5\textwidth]{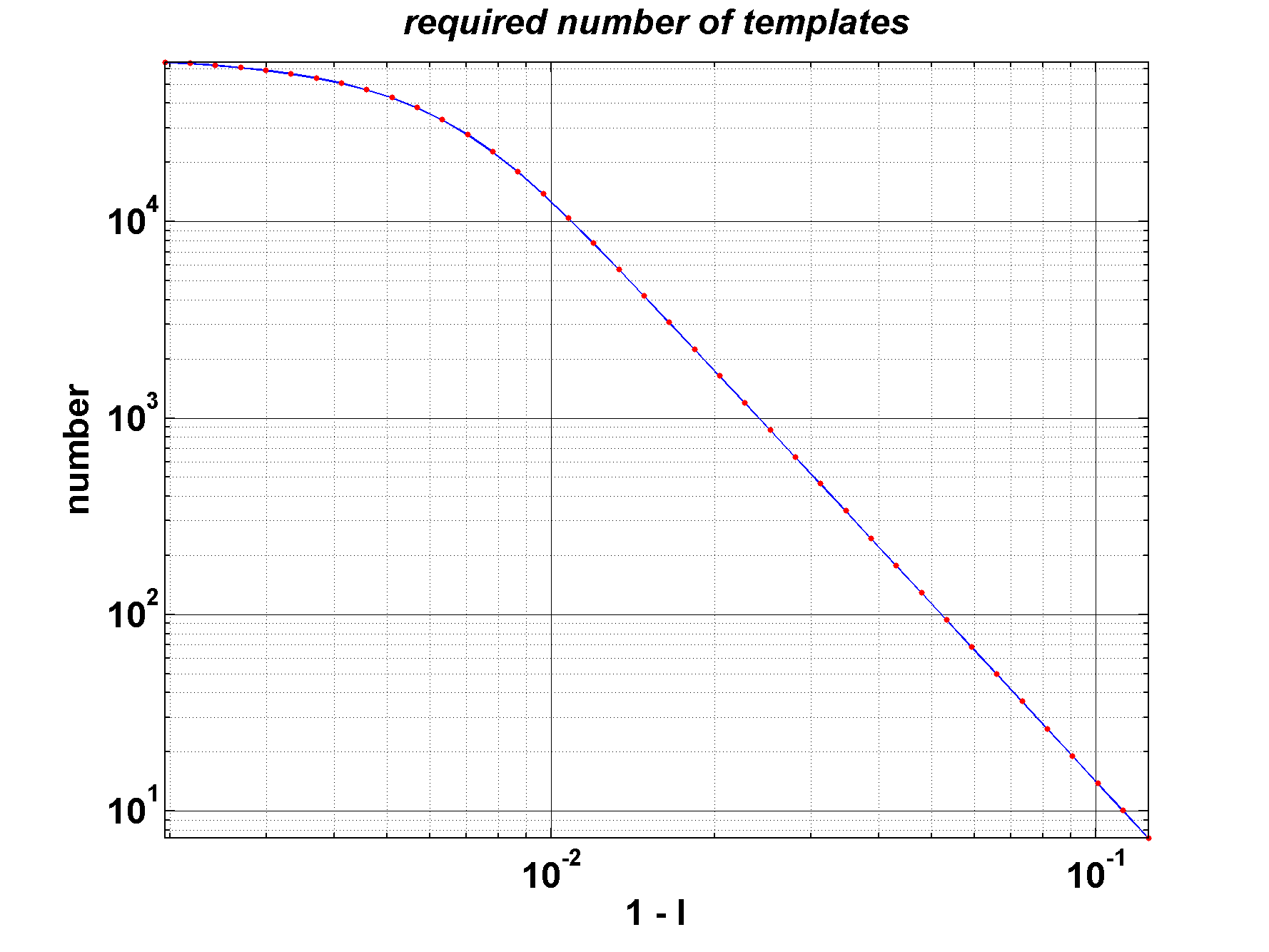}
\caption{ \label{fig:num} Number of templates required by the optimal
layout for different values of $l$. Notice that the original case
$l=1$ is not plotted here.}
\end{figure}

The structure of the likelihood surface can be
greatly simplified through the likelihood transform. Fig.~\ref{fig:likelihood3578}
(a)-(d) shows the several transformed likelihood surfaces $\mathcal{F}_{\bbsigma(l)}(\omega_0,\omega_1)$
with different values of $l$. When $l=3/4$, the modified likelihood surface
is very smooth. It is extremely simple to characterize the structure of this
likelihood surface or find its maximum. As $l$ increases, more and more
structures appear on the likelihood surface $\mathcal{F}_{\bbsigma(l)}(\omega_0,\omega_1)$.
It gradually converges to the original likelihood surface $\mathcal{F}(\omega_0,\omega_1)$.
These figures show how the difficulty of the optimization problem can be modified by the likelihood transform.
More precisely, we have calculated
the required number of templates for different values of $l$ in Fig.~\ref{fig:num}.

For $1-l>0.1$, the $0.03$-mismatch rule gives an error rectangular, which is
comparable to the area of the entire search parameter space. However, the error
rectangular may have very different shape from the search parameter space.
Therefore, in $1-l>0.1$ region, the required number of templates shown in
Fig.~\ref{fig:num} only serves as a rough estimate of the complexity of the modified
likelihood surface. In the more interesting $1-l<0.1$ region, the dependence of the
required number on $1-l$ roughly follows a power law. The required number decreases
rapidly in this region, hence the difficulty of search decreases rapidly.

These features of likelihood transform can potentially help the optimization
algorithms. For example, it may help in the design of efficient hierarchical
algorithms to search for GW signals.

In the following, we will show that in some cases the likelihood transform can even make deterministic optimization methods such as Newton's method possible. Hence, the search algorithm will be much more efficient.

In the neighbourhood of a (local) maximum $\bbtheta_{\bbsigma(l)}$ on modified likelihood surface $\mathcal{F}_{\bbsigma(l)}(\cdot)$, the geometry can be described by a Taylor series
\begin{eqnarray}\label{Eq:quadraticF}
\mathcal{F}_{\bbsigma(l)}(\bbtheta_{\bbsigma(l)}+\Delta\bbtheta)=\mathcal{F}_{\bbsigma(l)}(\bbtheta_{\bbsigma(l)})\nonumber \\
+ \frac{1}{2}\frac{\partial^2\mathcal{F}_{\bbsigma(l)}}{\partial \theta^\mu \partial \theta^\nu}\Delta\theta^\mu\Delta\theta^\nu
+ \mathcal{O}(\Delta\bbtheta^3).
\end{eqnarray}
where we have assumed the Einstein summation convention. The first derivative vanishes and the modified Fisher information matrix $I_{\mu\nu}^{\bbsigma(l)}=-\frac{\partial^2\mathcal{F}_{\bbsigma(l)}}{\partial \theta^\mu \partial \theta^\nu}\mid_{\bbtheta=\bbtheta_{\bbsigma(l)}}$ is positive definite due to the fact that $\mathcal{F}(\bbtheta_{\bbsigma(l)})$ is a maximum stationary point. Notice that, when $l=l_*$, the Taylor expansion is on the original likelihood surface around the best estimate $\bbtheta_*$, and $I_{\mu\nu}=I_{\mu\nu}^{\bbsigma_(l*)}=-\frac{\partial^2\mathcal{F}}{\partial \theta^\mu \partial \theta^\nu}\mid_{\bbtheta=\bbtheta_*}$ is the Fisher information matrix at the best estimate $\bbtheta_*$. For each modified likelihood surface $\mathcal{F}_{\bbsigma(l)}(\cdot)$, there exists a neighbourhood $\mathcal{B}_l\subseteq \mathcal{P}$ of $\bbtheta_{\bbsigma(l)}$ where the geometry of the likelihood surface can be approximated by a quadratic form quite well (say, the percentage error caused by higher order term is less $1\%$). According to our design, the smaller the $l$ the smoother the modified likelihood surface $\mathcal{F}_{\bbsigma(l)}(\cdot)$, hence the larger the neighbourhood $\mathcal{B}_l$. Sometimes, $\mathcal{B}_0$ can be as large as the entire parameter space $\mathcal{P}$.

Starting from any point $\bbtheta'_l\in \mathcal{B}_l$ on modified likelihood surface $\mathcal{F}_{\bbsigma(l)}(\cdot)$, one can easily find the best estimate $\bbtheta_{\bbsigma(l)}$ via some deterministic local-search algorithms. For instance, by neglecting higher order terms in Eq.~\ref{Eq:quadraticF} and differentiating both sides with respect to $\theta^\nu$, we get
\begin{eqnarray}
\frac{\partial\mathcal{F}_{\bbsigma(l)}(\bbtheta'_l)}{\partial \theta^\nu}=
\frac{\partial^2\mathcal{F}_{\bbsigma(l)}}{\partial \theta^\mu \partial \theta^\nu}\Delta\theta^\mu.
\end{eqnarray}
Thus, we calculate the best estimate in just one step
\begin{eqnarray}\label{Eq:iteration}
\theta_{\bbsigma(l)}^\mu&=&\theta'^\mu_l-\Delta\theta^\mu\nonumber\\
&=&
\theta'^\mu_l-\left[\frac{\partial^2\mathcal{F}_{\bbsigma(l)}}{\partial \theta^\mu \partial \theta^\nu}\right]^{-1}\frac{\partial\mathcal{F}_{\bbsigma(l)}(\bbtheta'_l)}{\partial \theta^\nu},
\end{eqnarray}
where $\frac{\partial^2\mathcal{F}_{\bbsigma(l)}}{\partial \theta^\mu \partial \theta^\nu}$ is constant in $\mathcal{B}_l$, so it can be calculated at $\bbtheta'_l$. Observe that as $l$ gradually runs from $0$ to $l_*$, $\mathcal{B}_l$ shrinks smoothly. Also, since $\mathcal{B}_l$ is roughly a quadratic region, $\bbsigma_l$ should be near the center of $\mathcal{B}_l$. So, there must exist a smaller region $\mathcal{B}_{l_1}$ (with $l_1>l$) which contains $\bbtheta_{\bbsigma(l)}$ in it. One can take $\bbtheta_{\bbsigma(l)}$ as the starting point in $\mathcal{B}_{l_1}$ and repeat Eq.~\ref{Eq:iteration} to calculate the best estimate $\bbtheta_{\bbsigma(l_1)}$ on $\mathcal{F}_{\bbsigma(l_1)}(\cdot)$. By iterating the above process, one will find the best estimate $\bbtheta_*$ on the original likelihood surface $\mathcal{F}(\cdot)$.

\begin{figure}
\includegraphics[width=0.5\textwidth]{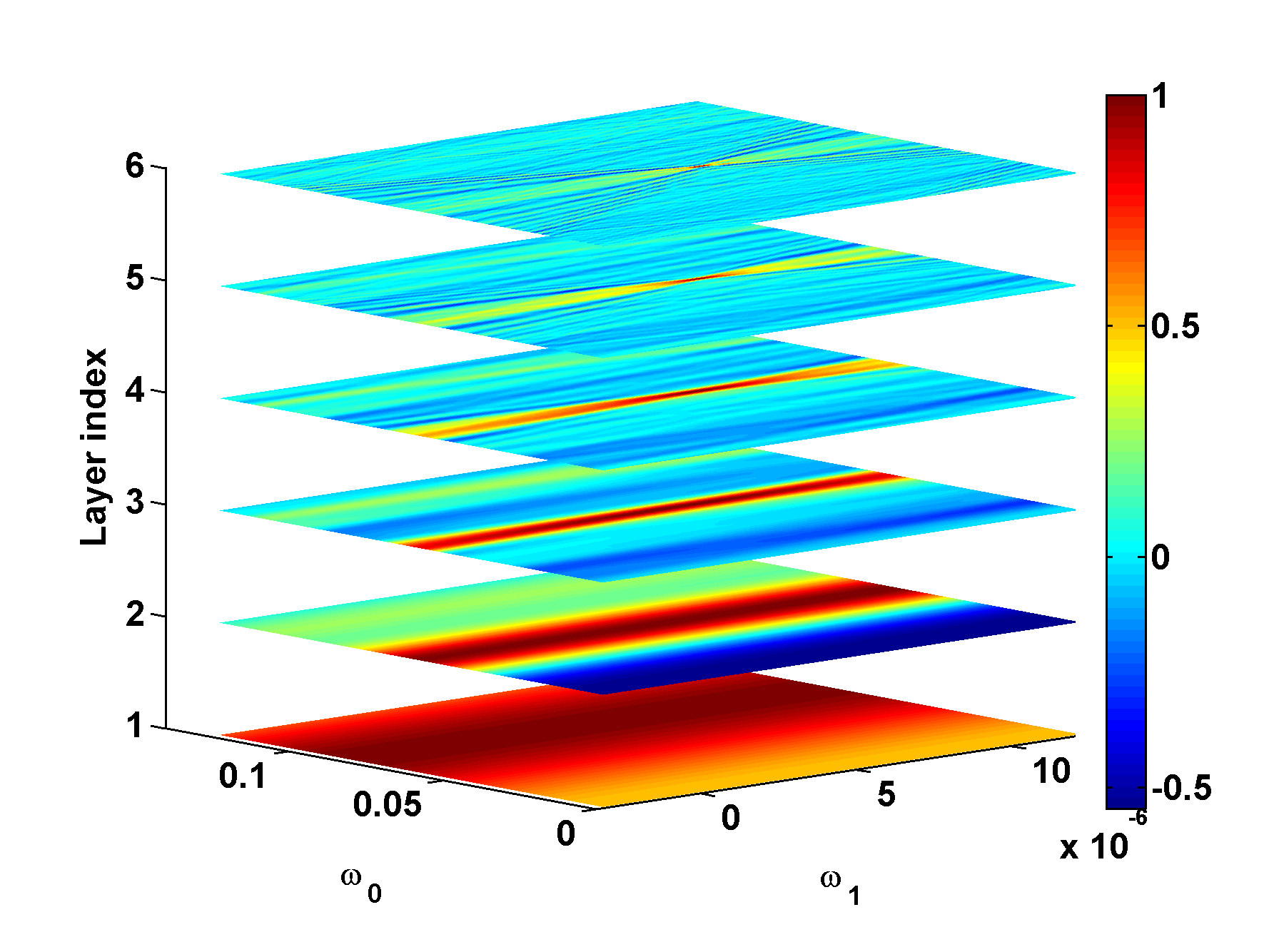}
\caption{ \label{fig:6layer} A plot of six transformed likelihood surfaces $\mathcal{F}_{\bbsigma (l)}$.}
\end{figure}

\begin{figure}[htbp]
\subfloat[]{
\begin{minipage}[t]{0.25\textwidth}
\centering
\includegraphics[width=1.0\textwidth]{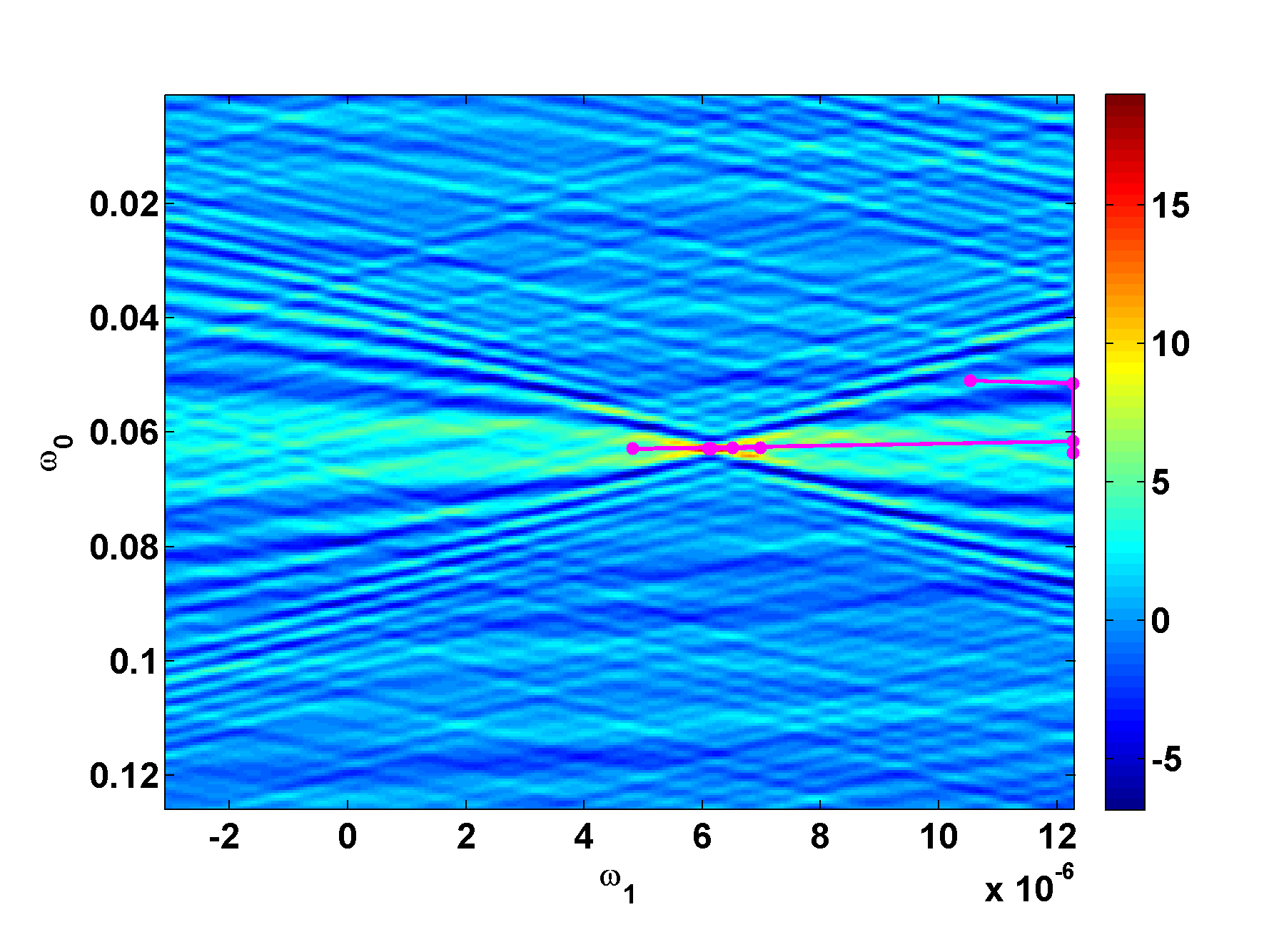}
\end{minipage}
}
\subfloat[]{
\begin{minipage}[t]{0.25\textwidth}
\centering
\includegraphics[width=1.0\textwidth]{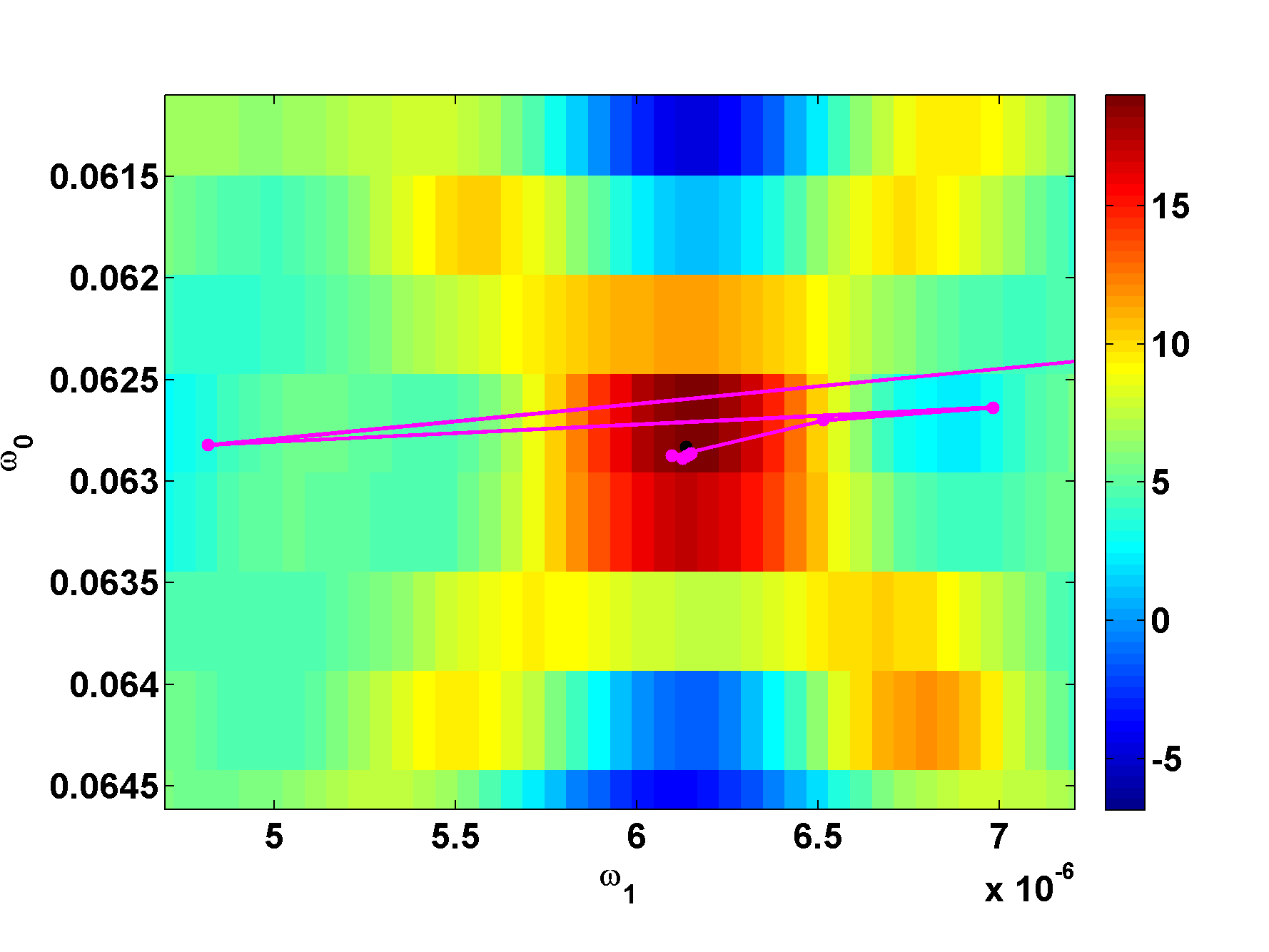}
\end{minipage}
}
\caption{ \label{fig:LTsearch} A deterministic search with the help of likelihood transform. The pink points identify the trajectory of $\theta_{\bbsigma(l)}^\mu$.
 (b). is a zoom-in version of (a). }
\end{figure}

Usually, we need to study the properties of the neighbourhood $\mathcal{B}_l$ in order to design an efficient deterministic algorithm. However, in some cases, likelihood
transform can change the likelihood surface to be so smooth and regular that we can simply choose a set of $\mathcal{Y}_\bbsigma$ to perform a deterministic search.
As an example, we still use the waveform model introduced in the last subsection and set the SNR to 20. Six transformed likelihood surfaces are shown in Fig.~\ref{fig:6layer}. Notice that the global maxima on these surfaces are normalized to 1. We start from 10 points in the parameter space uniformly sampled
in $\omega_0$ with random $\omega_1$. Then, we calculate the values of $\mathcal{F}_\bbsigma (\bbtheta)$ at these 10 points on the smoothest transformed likelihood
surface. The maximum among these 10 points is set as the initial location for the Newton's method with 10 iterations. Fig.~\ref{fig:LTsearch} shows the simulation
result of this deterministic algorithm. After 7 iterations, this algorithm converges to the location of the global maximum of the original likelihood surface. In this
process, we have only used a few tens of templates. Comparing to 69,620 templates required by a grid-based search algorithm, the deterministic algorithm is about 1,000
times more efficient.

\emph{Conclusion}--We have introduced the likelihood transform as a general tool to make optimization and parameter estimation easier.
The likelihood transform can gradually transform the likelihood surface to a smoother shape with less complex structure. On these
modified likelihood surfaces, the local and global maxima are much easier to find. Since these modified likelihood
surfaces are directly related to the original likelihood surface by the likelihood transform, one can find the global
maximum of the original likelihood surface more efficiently based on knowledge of the transformed likelihood surfaces.
We have shown the possibility to use likelihood transform to accelerate stochastic optimization methods. Compared to
simulated annealing, likelihood transform gives indications that it would accelerate the heuristics more efficiently.
We applied likelihood transform to a GW data analysis problem with a toy waveform model. Simulation results show that
likelihood transform can manipulate the structure of the original likelihood surface, hence allowing it to combine with
and accelerate a hierarchical search. We have also shown that for the toy waveform model with SNR$=20$, likelihood transform
make a deterministic search possible, which turns out to be 1,000 times more efficient than the exhaustive grid-based
search for GW signals. With the help of likelihood transform, a template-based deterministic search for GW signals is
shown to be possible for the first time.

In this article, we have only considered linear functionals, or more specifically, convolutions
with Gaussian kernels with uncorrelated covariances. In the future, we will study other linear functionals
and even nonlinear functionals $\mathcal{Y}_\bbsigma$, which would potentially exhibit better properties.

%%%%%%%%%%
\begin{acknowledgements}
The author would like to thank S. Babak, B. Krishnan and R. Prix for useful comments.
The author was partially supported by DFG Grant No. SFB/TR 7 Gravitational Wave
Astronomy. The author would also like to thank German Research Foundation for funding
the Cluster of Excellence QUEST-Center for Quantum Engineering and
Space-Time Research.
\end{acknowledgements}

%%%%%%%%%%%%%%%%%%%%%%%%%%%%%%%%%%%%%%%%%
\end{document}